\documentstyle[sprocl,epsfig]{article}

\def\photonatomrightt{\begin{picture}(3,1.5)(0,0)
                               \put(0,-0.75){\tencirc \symbol{2}}
                               \put(1.5,-0.75){\tencirc \symbol{1}}
                               \put(1.5,0.75){\tencirc \symbol{3}}
                               \put(3,0.75){\tencirc \symbol{0}}
                     \end{picture}
                    }

\def\photonrightthalf{\begin{picture}(15,1.5)(0,0)
                    \multiput(0,0)(3,0){5}{\photonatomrightt}
                 \end{picture}
                }

\def\fermionul{\begin{picture}(15,15)(0,0)
                        \put(0,0){\vector(-1,1){7.5}}
                        \put(-7.5,7.5){\line(-1,1){7.5}}
                  \end{picture}
                 }

\def\fermionur{\begin{picture}(15,15)(0,0)
                        \put(-15,-15){\vector(1,1){7.5}}
                        \put(-7.5,-7.5){\line(1,1){7.5}}
                  \end{picture}
                 }

\def\gaugebosonrighthalf{\begin{picture}(15,1)(0,0)
                            \put(0,0){\line(1,0){0.75}}
                            \multiput(2.25,0)(3,0){4}{\line(1,0){1.5}}
                            \put(14.25,0){\line(1,0){0.75}}
                         \end{picture}
                        }

\def\gaugebosonurhalf{\begin{picture}(15,15)(0,0)
                            \put(0,0){\line(1,1){15.0}}
                  \end{picture}
                 }

\def\gaugebosondrhalf{\begin{picture}(15,15)(0,0)
                            \put(0,0){\line(1,-1){15}}
                  \end{picture}
                 }

\newenvironment{Feynman}[3]{\begin{center}
                            \setlength{\unitlength}{#3 mm}
                            \begin{picture}(#1)(#2)
                            \thicklines
                           }{\end{picture} \end{center}}

\def\ra{\rightarrow}

\def\be{\begin{equation}}
\def\ee{\end{equation}}
\def\bea{\begin{eqnarray}}
\def\eea{\end{eqnarray}}

\def\ee{\mbox{$\rm e^+e^-$}}
\def\bb{\mbox{$\rm b\bar{b}$}}
\def\bbA{\mbox{$\rm \bb A$}}

\begin{document}

\topmargin = 0cm

{

\begin{titlepage}

\begin{center}
\mbox{ } 

\vspace*{-4cm}


\end{center}
\vskip 0.5cm
\begin{flushright}
\Large
\mbox{\hspace{10.2cm} hep-ph/9911344} \\
\mbox{\hspace{10.2cm} IEKP-KA/99-19} \\
\mbox{\hspace{11.5cm} Nov. 1999}
\end{flushright}
\Large
\begin{center}
\vskip 3cm
{\Large\bf
A DIRECT MEASUREMENT OF \\
\boldmath$\tan\beta$\unboldmath: 
\boldmath$ \ee\ra\bb\ra\bbA$\unboldmath\ AT A \\
FUTURE \boldmath$\rm e^+e^-$\unboldmath\ LINEAR COLLIDER}

\vskip 1cm
{\Large\bf 
M. Berggren$^a$, 
R. Ker\"anen$^b$,
A. Sopczak$^b$}

{\Large
\vspace*{2cm}
$^a$~LPNHE, Universit\'e de Paris VI \& VII

\vspace*{0.5cm}
$^b$~Karlsruhe University}
\end{center}

\vskip 1.8cm
\centerline{\Large \bf Abstract}

\vspace*{2cm}
\hspace*{-3cm}
\begin{picture}(0.001,0.001)(0,0)
\put(,0){
\begin{minipage}{16cm}
\Large
\renewcommand{\baselinestretch} {1.2}
The experimental sensitivity of the reaction $\ee\ra\bb\ra\bbA$
has been studied with a full-statistics background simulation for 
$\sqrt{s}=500$~GeV and ${\cal L}=500$~fb$^{-1}$. 
The simulation is based on a fast and realistic simulation of a
TESLA detector. For the first time this reaction has been analysed 
for a future linear collider and we show that a signal could be observed. 
A significant signal over background is achieved by the application of 
an Iterative Discriminant Analysis (IDA). For a signal production 
cross section of only 2~fb, which is expected for a Higgs boson mass 
of 100 GeV and $\tan\beta=50$, we achieve 100 signal over 100 background 
events, and obtain for a $\tan\beta$ measurement:
$\Delta\tan\beta / \tan\beta = 0.07$. This measurement
requires a high-luminosity future collider as proposed in the TESLA
project.
\renewcommand{\baselinestretch} {1.}

\normalsize 
\vspace{2cm}
\begin{center}
{\large \em
Talk at the Worldwide Workshop on Future $e^+e^-$ Collider, 
April 1999, Sitges, Spain, \\
to be published in the proceedings.
\vspace*{-7cm}
}
\end{center}
\end{minipage}
}
\end{picture}
\vfill

\end{titlepage}

\newpage
\thispagestyle{empty}
\mbox{ }
\newpage
\setcounter{page}{1}
}

\title{
A DIRECT MEASUREMENT OF \\
\boldmath$\tan\beta$\unboldmath: 
\boldmath$ \ee\ra\bb\ra\bbA$\unboldmath\ AT A \\
FUTURE \boldmath$\rm e^+e^-$\unboldmath\ LINEAR COLLIDER}

\author{R. KERANEN, A. SOPCZAK\footnote{speaker}}

\address{Karlsruhe University}

\author{M. BERGGREN}

\address{LPNHE, Universit\'e de Paris VI \& VII}


\maketitle\abstracts{
\vspace*{0.5cm}
The experimental sensitivity of the reaction $\ee\ra\bb\ra\bbA$
has been studied with a full-statistics background simulation for 
$\sqrt{s}=500$~GeV and ${\cal L}=500$~fb$^{-1}$. 
The simulation is based on a fast and realistic simulation of a
TESLA detector. For the first time this reaction has been analysed 
for a future linear collider and we show that a signal could be observed. 
A significant signal over background is achieved by the application of 
an Iterative Discriminant Analysis (IDA). For a signal production 
cross section of only 2~fb, which is expected for a Higgs boson mass 
of 100 GeV and $\tan\beta=50$, we achieve 100 signal over 100 background 
events, and obtain for a $\tan\beta$ measurement:
$\Delta\tan\beta / \tan\beta = 0.07$. This measurement
requires a high-luminosity future collider as proposed in the TESLA
project.}

\vspace*{0.5cm}
\section{Introduction}
\begin{figure}[tp]
\vspace*{-3mm}
\caption{\label{fig:feynman} Signal production process and expected event rate.}
\begin{minipage}{0.49\textwidth}
\vspace*{-1cm}
\begin{center}
\begin{Feynman}{65,40}{5,27}{0.9}
\put(25,40){\fermionul}        \put(5,22){${\rm e^-}$}
\put(25,40){\fermionur}        \put(5,55){${\rm e^+}$}
\put(15,30){\photonrightthalf} \put(32,30){$\gamma$}
\put(25,40){\gaugebosonrighthalf}  \put(30,43){${\rm Z^*}$}
\put(40,40){\gaugebosonurhalf} \put(57,55){${\rm b}$}
\put(40,40){\gaugebosondrhalf} \put(57,22){${\rm \bar{b}}$}
\put(50,50){\gaugebosondrhalf} \put(65,32){$\rm A $}
\end{Feynman}
\end{center}
\end{minipage}
\hfill
\begin{minipage}{0.49\textwidth}
\vspace*{2mm}
\begin{center}
\mbox{\epsfig{file=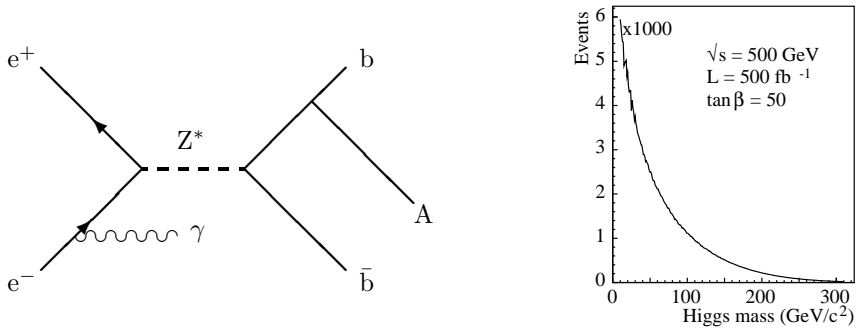,width=0.6\textwidth}}
\end{center}
\end{minipage}
\end{figure}

A future linear collider has a great potential for discovering
new particles and measuring their properties.
While many parameters of theories beyond the Standard Model
can be measured with high precision, the determination of the important
$\tan\beta$ parameter, the ratio of the vacuum expectation values 
of two Higgs boson doublets, is difficult, 
in particular when $\tan\beta$ is large.
This study uses the fact that the Yukawa coupling \bbA\ 
is proportional to $\tan\beta$. Therefore, the value of $\tan\beta$
can be directly derived from the measurement of the 
$\ee\ra\bb\ra\bbA\ra\bb\bb$ production cross section. The challenge
of this study is the low expected production rate and the large 
irreducible background for a four-jet final state.
The signal production process and the expected event rate~\cite{maria} 
are shown in Fig.~\ref{fig:feynman}. Searches for this four-jet channel
were performed at LEP with data taken at the Z resonance~\cite{l3,aleph,delphi}.

\section{Event Simulation}
The simulated production process is $\ee\ra\bb\ra\bbA\ra\bb\bb$,
leading to four b-quark jets in the final state.
The signal~\cite{generator} and background~\cite{background}
event generators 
include initial state radiation and beamstrahlung.
We simulated a 100 GeV pseudoscalar Higgs boson.
The generated events are passed through the fast detector
simulation SGV~\cite{sgv}.
This program follows the helical paths of charged particles
through the detector in order to estimate their track-parameter
covariance matrices.
The track-parameters are then smeared according to these matrices.
Calorimeters are simulated by parametrization of the
detector response function.
A probability-based b-tagging method is implemented using
the simulated track-parameters (measured mainly with the
micro-vertex detector).
Particle identification and track-finding efficiencies 
are also simulated.
The detector properties closely follow the TESLA detector CDR~\cite{tesla}.

\section{Event Preselection}
First, an event preselection is applied using the hadronic 
character and expected b-quark characteristics of the simulated signal. 
The following preselection cuts are applied: 3rd significant jet b-tag $ > 2$,
$N_{\rm cluster} > 17$,
$E_{\rm vis} / \sqrt{s} > 0.6$,
$E_{\rm neutral} / \sqrt{s} < 0.5$,
$E_{\gamma} < 30$~GeV,
thrust $ < 0.92$.
The number of simulated events for the signal and for each background 
channel, as well as the remaining events after the preselection are given in 
Table~\ref{tab:pre}.

\begin{table}[hp]
\caption{\label{tab:pre} Number of simulated signal and background events before and 
                         after the preselection.}
\begin{center}
\begin{tabular}{|c|c|c|c|c|c|c|c|c|}\hline
Channel      & bbA & qq  & WW   &eW$\nu$& tt  & ZZ  & eeZ  & hA \\ \hline
(in 1000)    & 50  &6250 & 3500 & 2500  & 350 & 300 & 3000 & 50 \\ \hline
After presel.& 73\% &20991 &7481&0 &89983&10278 & 145 & 12665   \\ \hline
%

\hline
\end{tabular}
\end{center}
\end{table}

\section{Iterative Discriminant Analysis}
\begin{figure}[hp]
\caption{\label{fig:btag} b-tagging 
variable and Fox-Wolfram moment after the preselection.}
\begin{center}
\vspace*{-0.7cm}
\mbox{\epsfig{file=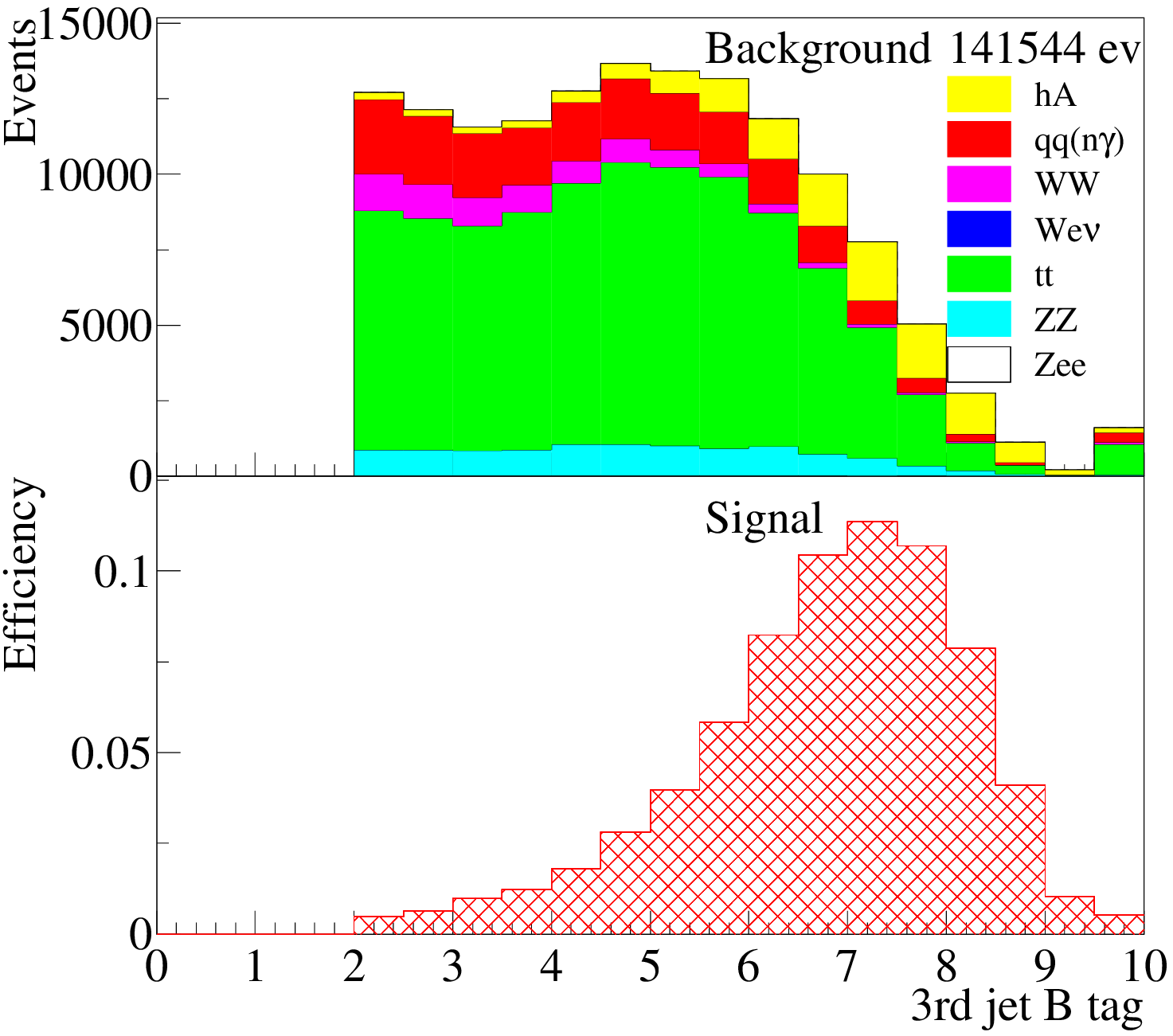,width=0.49\textwidth}}
\mbox{\epsfig{file=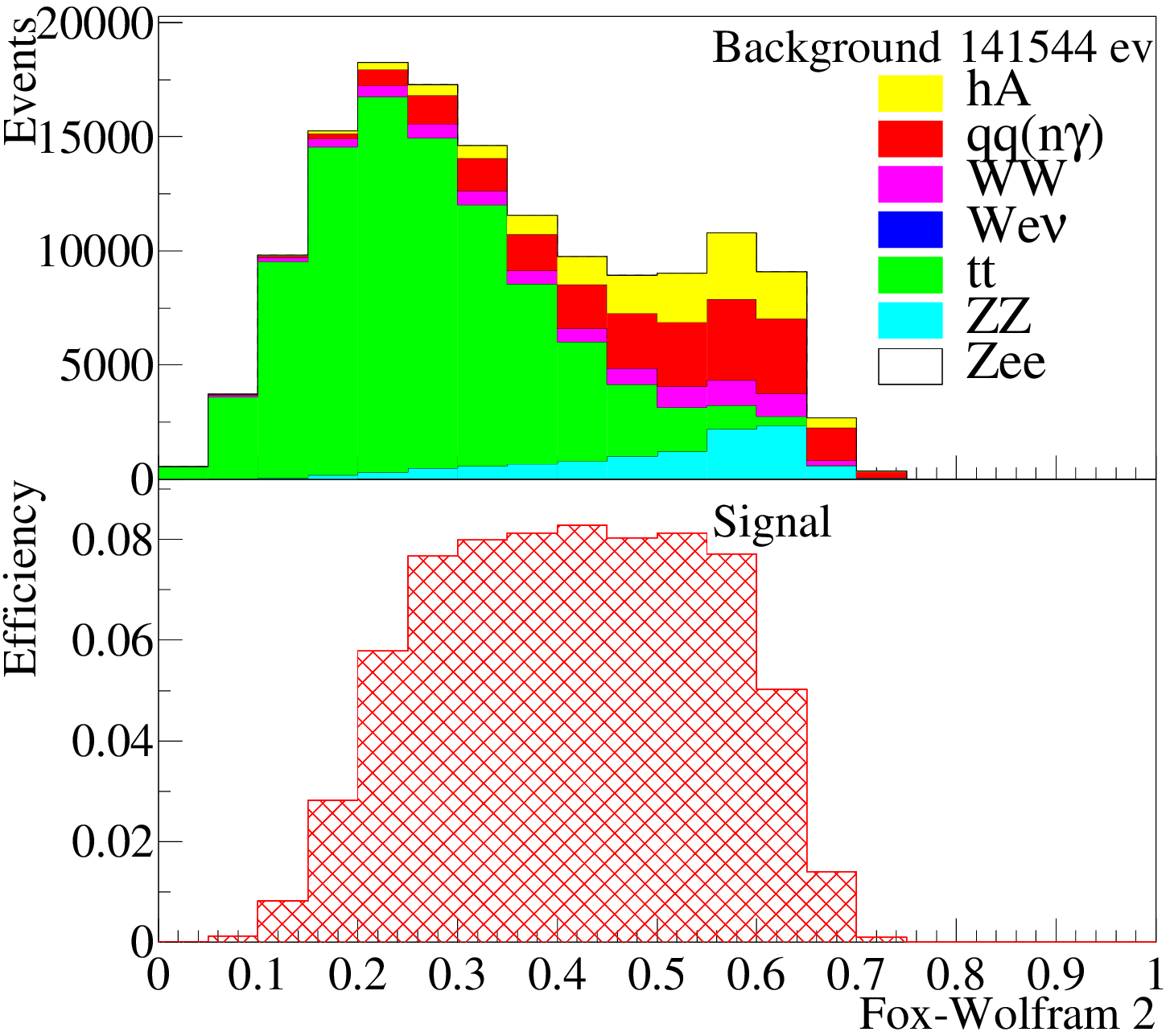,width=0.49\textwidth}}
\end{center}
\vspace*{-0.2cm}
\end{figure}

\begin{figure}[hp]
\vspace*{-5mm}
\caption{\label{fig:ida1} Thrust value and first step IDA output.}
\begin{center}
\vspace*{-0.7cm}
\mbox{\epsfig{file=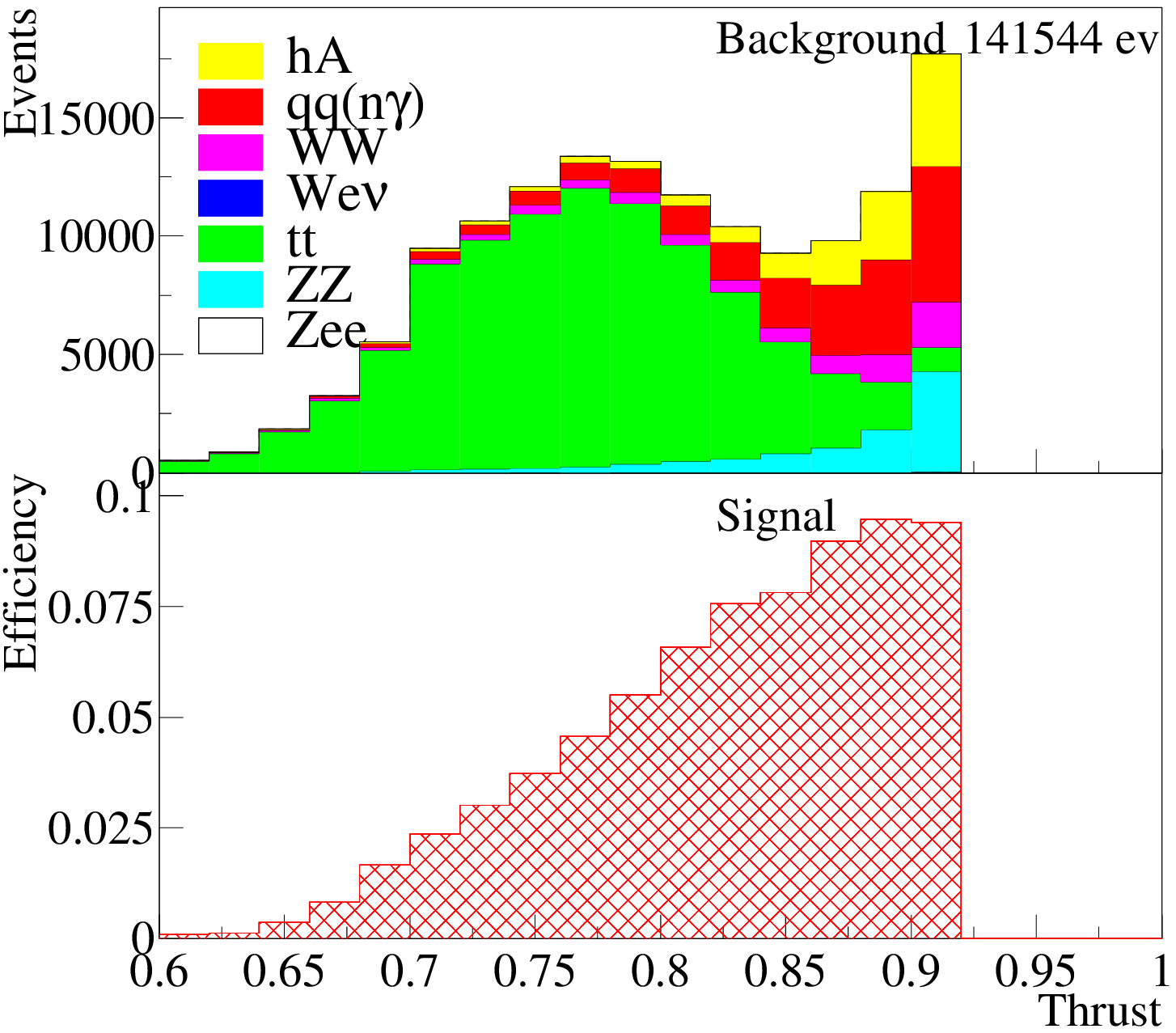,width=0.49\textwidth}}
\mbox{\epsfig{file=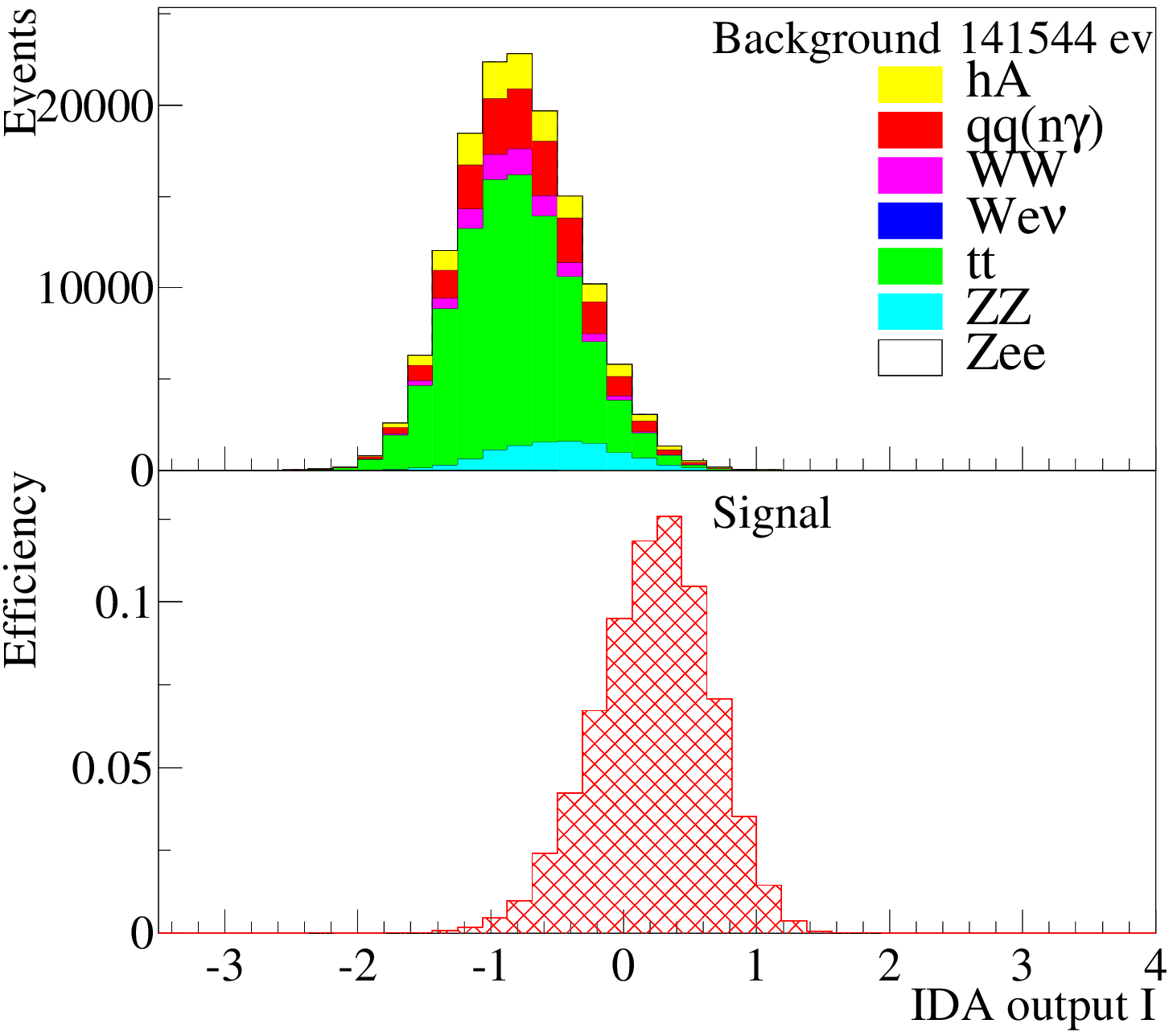,width=0.49\textwidth}}
\end{center}
\vspace*{-0.9cm}
\end{figure}

\begin{figure}[tp]
\caption{\label{fig:ida2} Final IDA output and background vs. signal 
efficiency.}
\begin{center}
\vspace*{-0.7cm}
\mbox{\epsfig{file=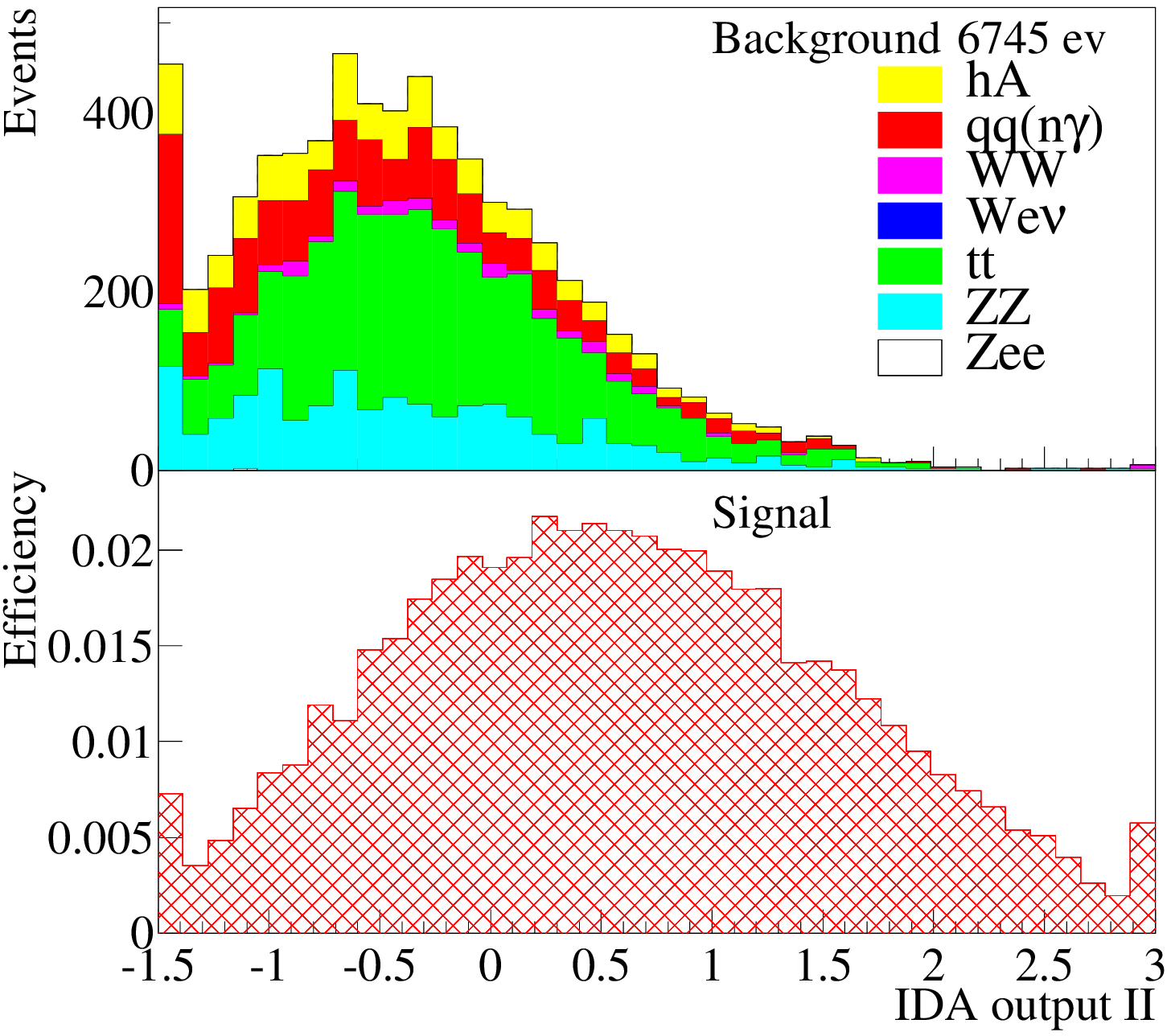,width=0.49\textwidth}}
\mbox{\epsfig{file=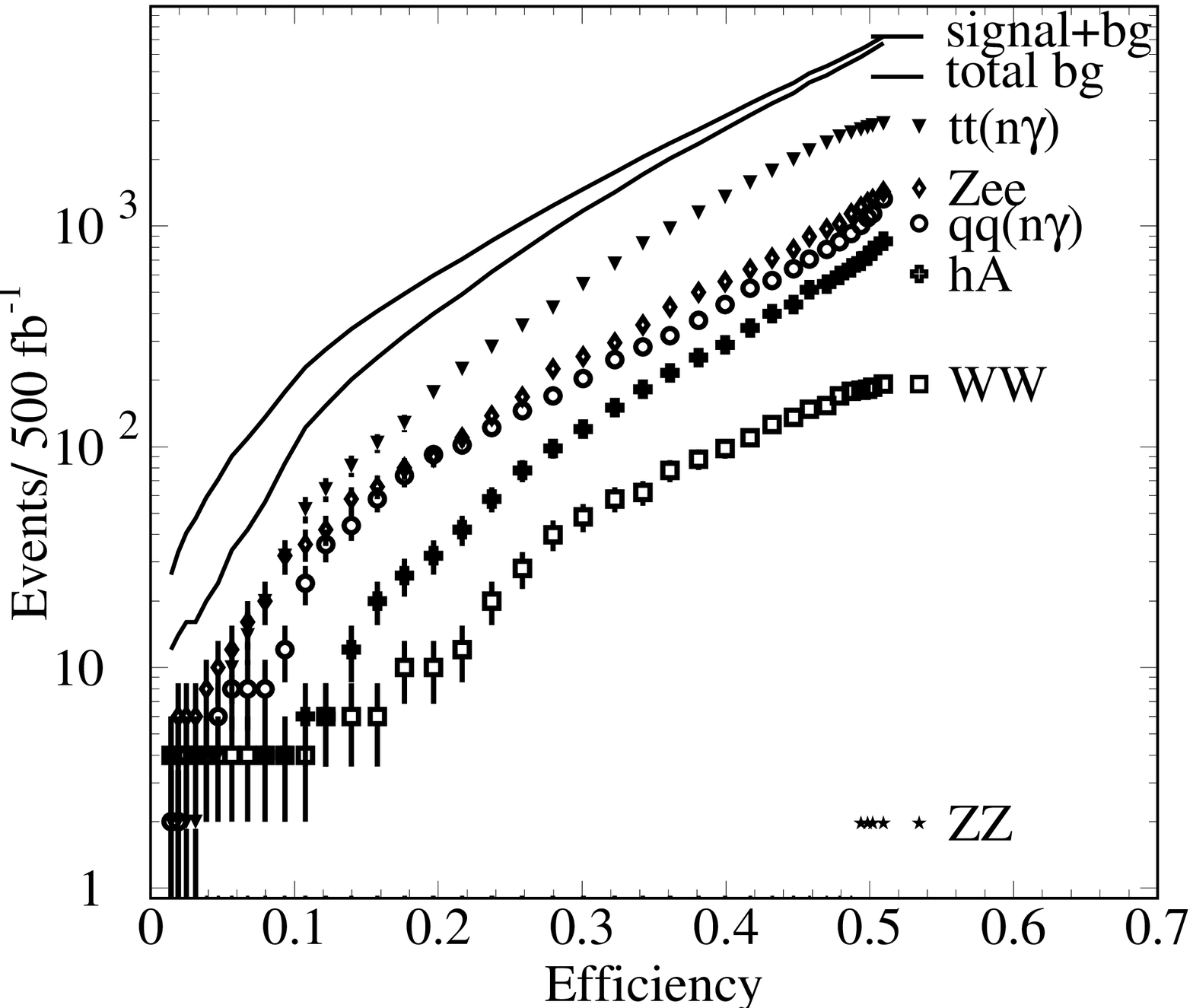,width=0.49\textwidth}}
\end{center}
\end{figure}

In order to separate the signal from the background, the following selection variables
are defined: highest and third significant jet b-tagging, thrust value,
second Fox-Wolfram moment, 
y-cut value to form four jets, minimum and maximum jet energy,
isolation angle of the most energetic jet, minimum angle and invariant mass between
any jet pair, minimum jet and event charge multiplicity, charged and neutral event
energy.
Figure~\ref{fig:btag} shows the simulated b-tagging variable for
the third significant jet and the second Fox-Wolfram moment after the preselection for
141544 remaining background events.
The thrust value and the IDA output variable are shown in Fig.~\ref{fig:ida1}.
Half of these events and half of the signal events are used to train the 
IDA~\cite{ida}.
In a first step, a cut on the IDA output variable is applied such that the 
efficiency is reduced by 30\%. The remaining signal and 6745 background events 
are again passed through the IDA. Figure~\ref{fig:ida2} shows the
IDA output variable and the resulting number of background events as a function 
of the signal efficiency.

\vspace*{0.5cm}
\section{Results}
We have determined the expected background rate for a given signal
efficiency and evaluated that the sensitivity for a 100 GeV 
pseudoscalar Higgs boson in the process $\ee\ra\bb\ra\bbA$ suffices to
determine the value of $\tan\beta$.
The sensitivity $N_{\rm signal} / \sqrt{N_{\rm background}}$
is almost independent of the working point signal efficiency 
in the range 5\% to 50\%.
For a working point of 10\% efficiency, the total simulated background of about 
16 million events is reduced to 100 background events.
The resulting error on $\tan\beta= 50$ is 7\%:
$$
\Delta\tan^2\beta / \tan^2\beta = \Delta N_{\rm signal} / N_{\rm signal}
=\sqrt{N_{\rm signal} + N_{\rm background}} / N_{\rm signal} =  0.14.
$$
For smaller values of $\tan\beta$ the sensitivity reduces quickly.
A $5\sigma$ signal detection is possible for $\tan\beta= 35$.

In conclusion, an IDA anlalysis based on experience at LEP2
was applied and gave sufficient sensitivity to detect the signal process. 
A high-luminosity linear collider is essential and unique 
for this channel and allows the value of $\tan\beta$ to be measured with 
precision.

\section*{References}

\end{document}